# Iterative phase retrieval in coherent diffractive imaging: practical issues

Tatiana Latychevskaia, tatiana@physik.uzh.ch

## ABSTRACT

In this work, issues in phase retrieval in the coherent diffractive imaging (CDI) technique, from discussion on parameters for setting up a CDI experiment to evaluation of the goodness of the final reconstruction, are discussed. The distribution of objects under study by CDI often cannot be cross-validated by another imaging technique. It is therefore important to make sure that the developed CDI procedure delivers an artifact-free object reconstruction. Critical issues that can lead to artifacts are presented and recipes on how to avoid them are provided.

## Contents

# 1. Introduction

Coherent diffractive imaging (CDI) [1] and associated iterative phase retrieval methods [2] have been successfully applied in the past two decades for optical, x-ray, and electron imaging. However, it is common knowledge that people who start to develop iterative phase retrieval algorithms for the analysis of acquired diffraction patterns often characterize the whole process as unreliable. Often the iterative phase retrieval fails not because of the algorithms but because of improper preparation of data for the iterative phase retrieval. The purpose of this work is to provide some basics of phase retrieval in CDI, highlight some typical issues and discuss their solutions.

# 2. Basics of CDI experiment
## 2.1. Diffraction pattern formation

An object distribution can be described by a complex-valued transmission function $o(\vec{r})$, where $\vec{r}=(x,y,z)$ is the coordinate in the object domain. A plane wave is incident on an object and the distribution of the scattered wave in the detector plane is calculated by the following integral transformation:

$$U(\vec{R}) = -\frac{i}{\lambda}\iiint \exp(ikz)\, o(\vec{r})\, \frac{\exp\left(ik\left|\vec{R}-\vec{r}\right|\right)}{\left|\vec{R}-\vec{r}\right|}\, d\vec{r}, \qquad (1)$$

where $\lambda$ is the wavelength of the employed probing wave, $k=\frac{2\pi}{\lambda}$ is the wavenumber, $\vec{R}=(X,Y,Z)$ is the coordinate in the detector plane, $\exp(ikz)$ is the incident plane wave, $\left|\vec{R}-\vec{r}\right|$ is the distance between a point in the object plane $P_0$ and a point in the detector plane $P_1$, as illustrated in Fig. 1, and the integration is performed over all scattering elements of the object.

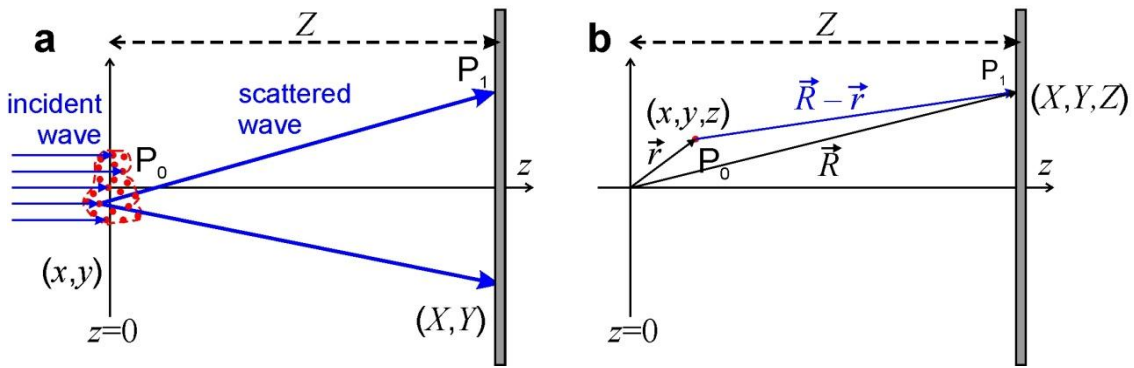

Fig. 1. Geometrical arrangement in coherent diffractive imaging. (a) General scheme. (b) Illustration to the vector definitions.

**Cartesian coordinates.** When the distance between the sample and the detector is much larger than characteristic length of the sample, so that $Z^2 \gg x_{max}^2 + y_{max}^2$, the following approximation can be applied:

$$\left|\vec{R}-\vec{r}\right| \approx (Z-z) + \frac{(x-X)^2+(y-Y)^2}{2Z}.$$

Assuming that $\exp\left[\frac{i\pi}{\lambda Z}(x^2+y^2)\right] \approx 1$ because of large $Z$, the wavefront distribution in the detector plane can be re-written as

$$U(X,Y) \approx -\frac{i}{\lambda Z}\exp(ikZ)\exp\left[\frac{i\pi}{\lambda Z}(X^2+Y^2)\right]\iint\left[\int o(x,y,z)dz\right]\exp\left[-\frac{i2\pi}{\lambda Z}(xX+yY)\right]dxdy, \quad (2)$$

which is a two-dimensional (2D) Fourier transform (FT) of the projected object distribution given by integration along the optical axis $o_z(x,y) = \int o(x,y,z)dz$.

In some situations the wavefront propagating through the sample experiences many scattering events. In this case the far-field distribution of the scattered wavefront is given 2D FT of the exit wave $U_0(x,y)$, which is a distribution immediately after the specimen:

$$U(X,Y) \approx -\frac{i}{\lambda Z}\exp(ikZ)\exp\left[\frac{i\pi}{\lambda Z}(X^2+Y^2)\right]\iint U_0(x,y)\exp\left[-\frac{i2\pi}{\lambda Z}(xX+yY)\right]dxdy. \quad (3)$$

**K-coordinates.** When the distance between the sample and the detector is much larger than the characteristic length of the sample, so that $R \gg r$, the following approximation can be applied:

$$\left|\vec{r}-\vec{R}\right| = \sqrt{R^2 - 2\vec{R}\vec{r} + r^2} \approx R - \frac{\vec{R}\vec{r}}{R}.$$

Next, we introduce *K*-coordinates as follows: $\vec{K} = \left(\frac{X}{R}, \frac{Y}{R}, \frac{Z}{R}\right)$, where $R = \sqrt{X^2+Y^2+Z^2}$ and $\left|\vec{K}\right| = K = 1$. The distribution of the wavefront in the detector plane according to Eq. 1 is given by:

$$U(\vec{K}) \approx -\frac{i}{\lambda R}\exp(ikR)\iiint \exp(ikz)o(\vec{r})\exp(-ik\vec{K}\vec{r})d\vec{r}. \quad (4)$$

Taking into account that $K_x^2 + K_y^2 + K_z^2 = 1$, this integral can be re-written as [3]

$$U(K_x, K_y) \approx -\frac{i}{\lambda R} \exp(ikR) \times$$

$$\times \iiint \exp(ikz) o(x,y,z) \exp\left[-ik\left(xK_x + yK_y\right)\right] \exp\left[-ikz\sqrt{1-K_x^2-K_y^2}\right] dxdydz.$$

Since the condition $K_x^2 + K_y^2 + K_z^2 = 1$ is an equation of a sphere in *K*-coordinates, an integral in the form of Eq. 4 can be applied for an optical system where the detector is spherical, or where the detected signal is in *K*-coordinates, as for example, in angular resolved photo-emission spectroscopy (ARPES) [4]. At a small numerical aperture $K_x^2 + K_y^2 \ll 1$, and the diffraction integral can be approximated to

$$U(K_x, K_y) \approx -\frac{i}{\lambda Z} \exp(ikR) \iint \left[\int o(x,y,z) dz\right] \exp\left[-ik\left(xK_x + yK_y\right)\right] dxdy, \qquad (5)$$

which is a 2D FT of the projected object distribution given by integration along the optical axis $o_z(x,y) = \int o(x,y,z) dz.$

Diffraction patterns can be transformed from Cartesian to K-coordinates and vice versa by applying the coordinate transformation procedure described elsewhere [5].

We just have demonstrated that the object distribution and the wavefront in the far-field constitute Fourier pair that is independent of the coordinate system. Therefore, in the following sections we will not specify the type of the coordinate systems (Cartesian or *K*-coordinates). A general transformation which describes the formation of a diffraction pattern in CDI is thus given by:

$$I(v,w) = |F(v,w)|^2 = \left|\iint f(x,y) \exp\left[-i(xv+yw)\right] dxdy\right|^2, \qquad (6)$$

where $f(x,y) = |f(x,y)| \exp[i\eta(x,y)]$ and $F(v,w) = |F(v,w)| \exp[i\psi(v,w)]$ are the complex-valued distributions in the object and Fourier domains, respectively, and $(v,w)$ are the coordinates in the Fourier domain.

A 2D object distribution is reconstructed from a 2D diffraction pattern. When sample consisting of individual scatterers, neglecting multiple scattering, the reconstructed 2D distribution is the projected object distribution (according to Eqs. 2 and 5), which does not depend on the *z*-extent of the scatterers distribution. For a sample where multiple scattering occurs, the reconstructed 2D distribution is the exit wave distribution (according to Eq. 3). The exit wave carries information about all the individual multiple scattering events and therefore its phase distribution is non-zero. In this case, the object constraints in the iterative phase retrieval routine should be designed as for phase objects.

## 2.2 Oversampling

If the complex-valued distribution of the scattered wavefront was available, then a simple inverse FT would give the sample distribution. In reality, only the intensity distribution can be measured, which provides only the amplitude distribution of the scattered wave. The missing phase distribution of the scattered wave, and with this the complete complex-valued distribution of the scattered wavefront, can however be recovered provided that the acquired intensity distribution is sufficiently fine-sampled ("oversampled"). The principle of "oversampling" of the diffraction pattern intensity is well-explained by Miao et al [6]. Here, we only briefly address some of the main points. The measured intensity provides the values of the magnitude of the Fourier transform:

$$|F(\vec{u})| = \left| \sum_{\vec{r}=0}^{N-1} f(\vec{r}) \exp(-i\vec{u} \cdot \vec{r}) \right|. \qquad (7)$$

Equation 7 is a set of equations where $f(\vec{r})$ are the unknowns. For an *n*-dimensional object, the total number of equations corresponds to the number of measured intensity pixels and is $N^n$. In the object domain the total number of pixels is $N^n$, where $M^n$ pixels have unknown values. For a complex-valued object sampled with $M^n$ pixels, the number of unknowns is $2M^n$. According to Friedel's law, a real-valued object has a diffraction pattern with symmetry $I(\vec{u}) = I(-\vec{u})$, and thus the number of equations is reduced to $N^n/2$ and the number of unknowns is $M^n$. The set of equations Eq. 7 can have a solution if the number of unknowns is less than the number of equations. This translates into conditions $N^n > 2M^n$ for complex-valued object and $N^n/2 > M^n$ for a real-valued object, respectively, which can be re-written as $(N/M)^n > 2$ or $N/M > 2^{1/n}$. The following ratio is introduced in the object domain [6]:

$$\sigma_0 = \frac{\text{total pixel number}}{\text{unknown-valued pixel number}}. \qquad (8)$$

Equations 7 should be in principle solvable if $\sigma_0 > 2$. Thus, to solve the phase problem, the magnitude of the Fourier transform should be oversampled to make the ratio $\sigma_0 > 2$. This leads to the oversampling condition which should be fulfilled in each dimension:

$$\sigma > 2^{1/n}, \qquad (9)$$

where $\sigma$ is the linear oversampling ratio, referred to as "oversampling ratio" in this study. Note that the required oversampling ratio is less for a three-dimensional (3D) object ($\sigma > 2^{1/3}$) than it is for a 2D object ($\sigma > 2^{1/2}$). It has been demonstrated that for 1D signals, no unique solution to the problem of recovering a signal from the amplitude of its Fourier transform exists [7].

The term "oversampling" originated from the fact that padding the object distribution (unknown-valued pixels) with known-valued pixels (as for example, padding by zeroes or zero-padding) in the object domain automatically leads to a finer sampling rate ("oversampling") of the signal spectrum in the Fourier domain.

## 2.3 Iterative phase-retrieval algorithms

Most iterative phase-retrieval algorithms are based on the Gerchberg-Saxton (GS) algorithm where two intensity measurements, in the object and the image planes, are utilized [8]. The phase distributions in these two planes are reconstructed by propagating the complex-valued wavefront between the two planes back and forth and replacing the updated amplitudes at each iteration with the measured amplitudes. In CDI, however, only one intensity measurement is available – the diffraction pattern, and some *a priori* information about the object distribution is known. As in the GS algorithm, the complex-valued wavefront is propagated back and forth between the object plane and the diffraction pattern (detector) plane, where the following constraints are applied. In the detector plane the amplitude of the updated wavefront is replaced by the square root of the measured intensity. In the object plane the object distribution must be surrounded by known values, typically zeros, in accordance with the oversampling condition discussed above. This constraint in combination with other requirements, as for example, that the object distribution must be real and positive, constitutes the object "support".

Two most popular iterative phase-retrieval algorithms are the error-reduction (ER) and the hybrid input-output (HIO) algorithms, which were introduced by Fienup [2]. We provide below some necessary details of these algorithms. The task is to recover the object complex-valued distribution $f(x,y)$ from the measured intensity distribution $I(v,w) = |F(v,w)|^2$. The reconstruction is done by applying an iterative procedure as depicted in Fig. 2.

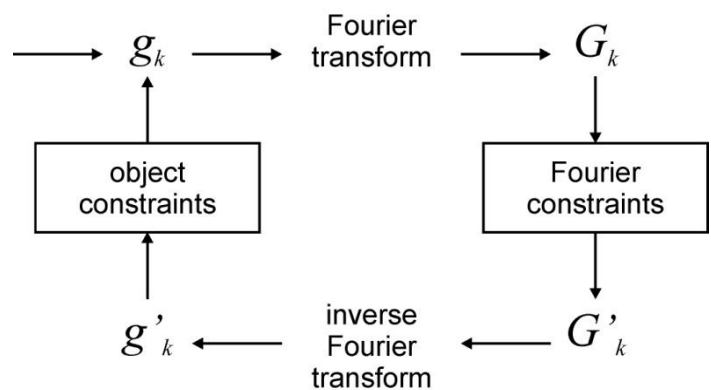

Fig. 2. Schematics of a general algorithm for iterative phase retrieval.

### 2.3.1 Error-reduction algorithm (ER)
The steps of the error-reduction (ER) algorithm at *k*-th iteration[2]:

(i) $G_k(v,w) = |G_k(v,w)| \exp[i\varphi_k(v,w)] = \mathcal{F}[g_k(x,y)],$

(ii) $G'_k(v,w) = |F(v,w)| \exp[i\varphi_k(v,w)],$

(iii) $g'_k(x,y) = |g'_k(x,y)| \exp[i\vartheta'_k(x,y)] = \mathcal{F}^{-1}[G'_k(v,w)],$ (10)

(iv) $g_{k+1}(x,y) = \begin{cases} g'_k(x,y), & \text{if } (x,y) \in \gamma, \\ 0, & \text{if } (x,y) \notin \gamma, \end{cases}$

where $g_k$, $\vartheta'_k$, $G'_k$, and $\varphi_k$ are estimates of $f$, $\eta$, $F$, and $\psi$, respectively, and $\mathcal{F}$ and $\mathcal{F}^{-1}$ denote FT and inverse FT, respectively. $\gamma$ is the set of points at which $g'_k(x,y)$ satisfies the object-domain constraints such that $g'_k(x,y)$ is positive real-valued (optional) and does not exceed the known diameter of the object (support constraint).

For the first iteration, the estimate of the object distribution $g_k(x,y)$ ($k$ = 1 for the first iteration) can be obtained by various approaches. For example: (1) A complex-valued far-field distribution is obtained by combining the square root of the measured intensity as the amplitude and a distribution of random numbers in the range from -π to π as the phase. The inverse FT of the obtained distribution gives $g_1(x,y)$. (2) A random complex-valued distribution multiplied with a known support gives $g_1(x,y)$.

The name "error-reduction" originates from the fact that during the iterative routine, the error at *k*-th iteration defined as $E^2_{Fk} = N^{-2} \sum_{u,w} \left[|G_k(v,w)| - |F(v,w)|\right]^2$ can only be the same or less than the error at the previous iteration as explained in detail elsewhere [2].

### 2.3.2 Hybrid input-output algorithm (HIO)
The hybrid input-output algorithm (HIO) [2] is obtained from ER algorithm by modifying the constraint in the object domain. The steps of the HIO algorithm at *k*-th iteration are the same as for ER algorithm except for step (iv):

(iv) $g_{k+1}(x,y) = \begin{cases} g'_k(x,y), & \text{if } (x,y) \in \gamma, \\ g_k(x,y) - \beta g'_k(x,y), & \text{if } (x,y) \notin \gamma, \end{cases}$ (11)

where $\beta$ is a constant. Typically, $\beta = 0.9$ is selected. The meaning of $\beta$ can be explained by analogy with the constant factor of the linear component $K_\text{P}$ in a proportional–integral–derivative (PID) controller. When $K_\text{P} \ll 1$, the signal very slowly converges to the desired value. When $K_\text{P} > 1$ the signal exhibits large oscillations and slow convergence to the desired value. When $K_\text{P}$ is selected to be less just below 1, the signal smoothly approaches to the desired value.

### 2.3.3 Shrinkwrap algorithm

The shrinkwrap algorithm [9] is a further modification of the HIO algorithm, where the object support in step (iv) is re-adjusted during the iterative reconstruction at each 20-th iteration. Eventually, the object support approaches the exact shape of the object distribution, thus providing a "tight support".

A recent review on various phase retrieval algorithms is provided by Shechtman et al [10].

## 2.4 Error functions

The convergence of the iterative reconstruction process and the goodness of the reconstruction can be monitored by an error function. For the ER-based algorithms the error function evaluates how well the iteratively recovered amplitudes match the measured amplitudes in the detector plane [2]:

$$\text{Error}\_F_k = \left\{ \frac{N^{-2} \sum_{u,w} \left[ |G_k(v,w)| - |F(v,w)| \right]^2}{\sum_{u,w} |F(v,w)|^2} \right\}^{1/2}. \qquad (12)$$

For the HIO-based algorithms the error functions evaluates how well the recovered object distribution satisfies the object constraints:

$$\text{Error}\_f_k = \left\{ \frac{\sum_{x,y \notin \gamma} |g'_k(x,y)|^2}{\sum_{x,y \in \gamma} |g'_k(x,y)|^2} \right\}^{1/2}. \qquad (13)$$

This type of error function is often employed in Miao et al works [6, 11, 12].

# 3. Practical issues
## 3.1 Setting up experiment

In this section we consider the issues which can (and should) be taken care of during the preparation of a CDI experiment. Previously, Thibault and Rankenburg provided a tutorial on how to perform a light optical CDI experiment in a teaching laboratory [13], however, their sample contained a reference scatterer, which thus created a Fourier transform holography (FTH) scheme [14, 15] - a technique which is different from CDI and requires a much easier reconstruction procedure.

### 3.1.1 Fulfilling the oversampling condition

Although one might often read that object was reconstructed from its diffraction pattern without any *a priori* information, this is not entirely correct. Some information about the object, namely the extent of the object in each dimension, must be known even before the experimental setup for conventional CDI is designed. The parameters of the experimental setup, namely: the extent of the object, the wavelength, the detector size and the number of pixels for sampling the diffraction pattern, must be selected in such way that the oversampling condition given by Eq. 9 is fulfilled.

An iterative reconstruction routine typically employs a fast Fourier transform (FFT), which performs a digital Fourier transform (DFT) in an optimized way. For 1D signals, the DFT is given by:

$$F_q = \sum_{p=0}^{N-1} f_p \exp(-2\pi i q p / N),$$

which when compared to Eq. 7 gives the relation between the pixel size in the object and the Fourier domains:

$$\Delta_q \Delta_p = \frac{2\pi}{N}.$$

This gives the side-length of the reconstructed object area $S = N\Delta_x = \frac{2\pi}{N\Delta_v}$. From Eq. 9, the oversampling ratio is given by: $\sigma = \frac{S}{S_0} = \frac{2\pi}{N\Delta_v S_0}$, where $S_0$ is the extent of the object itself, which must be selected in such a way that the oversampling condition provided by Eq. 9 is satisfied. The wavelength of the employed waves is accounted in $\Delta_v$. For Cartesian coordinates $\Delta_v = \frac{2\pi}{\lambda Z}\Delta_D$, where $\Delta_D$ is the detector pixel size, and $\Delta_v = \Delta_k = \frac{2\pi}{\lambda}\Delta_K$ for *K*-coordinates, where $\Delta_K$ is the pixel size in *K*-coordinates.

### 3.1.2 Different oversampling ratio

The oversampling condition should be fulfilled in each dimension. Figure 3a – b illustrate a faulty reconstruction of a 2D object when the oversampling condition was fulfilled only in one dimension. Figure 3c – f illustrate that once the oversampling condition is fulfilled, the sample distributions can be various: the sample must not be localized in one place, its parts can be scattered over the entire imaged area. Figure 4 demonstrates that speed of convergence or the error does not depend on the oversampling ratio.

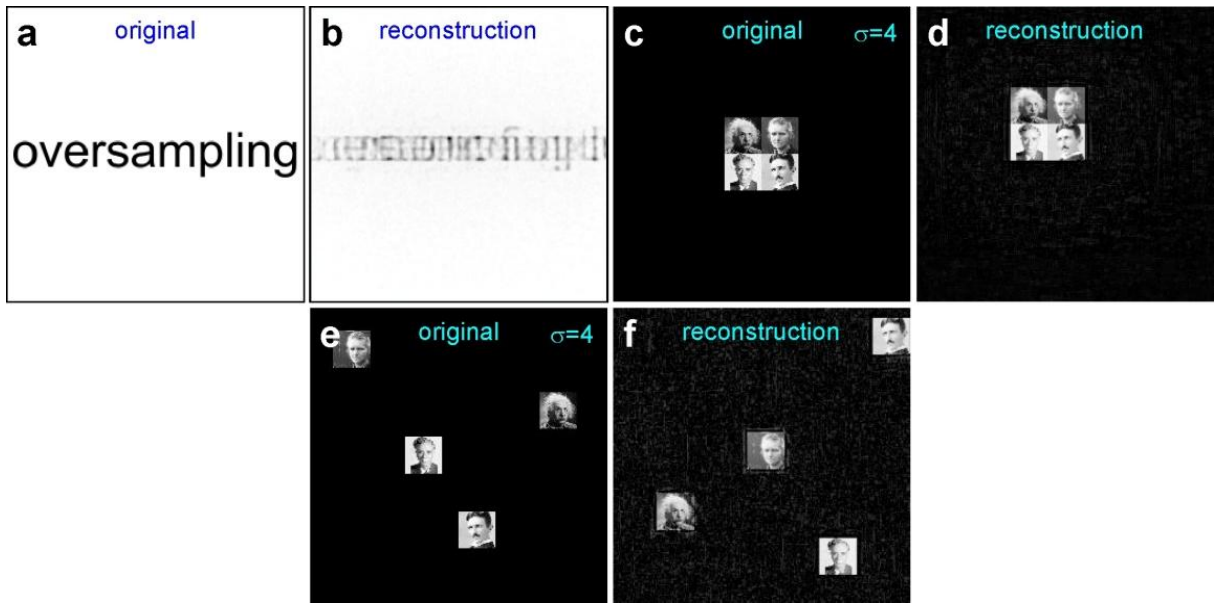

Fig. 3. Effects of oversampling ratio.

(a) Sample distribution and (b) the reconstruction obtained from its diffraction pattern where the oversampling condition was not fulfilled along $K_x$-dimension. The reconstruction was obtained by applying the shrinkwrap algorithm for 2000 iterations, 10 reconstructions with the smallest errors were selected, aligned and averaged; the number of pixels was 256 × 256 pixels.

(c) Sample distribution consisting of four objects placed together in the center and (d) its reconstruction obtained from the diffraction pattern with the parameters as described in Appendix 1.

(e) Sample distribution consisting of four objects scattered over the entire imaged area and (f) reconstruction obtained from its diffraction pattern. The oversampling ratio was $\sigma = 4$. The reconstruction was obtained by applying the shrinkwrap algorithm for 2000 iterations, one reconstruction with the least error was selected; the number of pixels was 256 × 256 pixels. Note that because the shrinkwrap algorithm begins with obtaining support from cross-correlation of the object, one part of the sample will always show up in the center of the reconstruction.

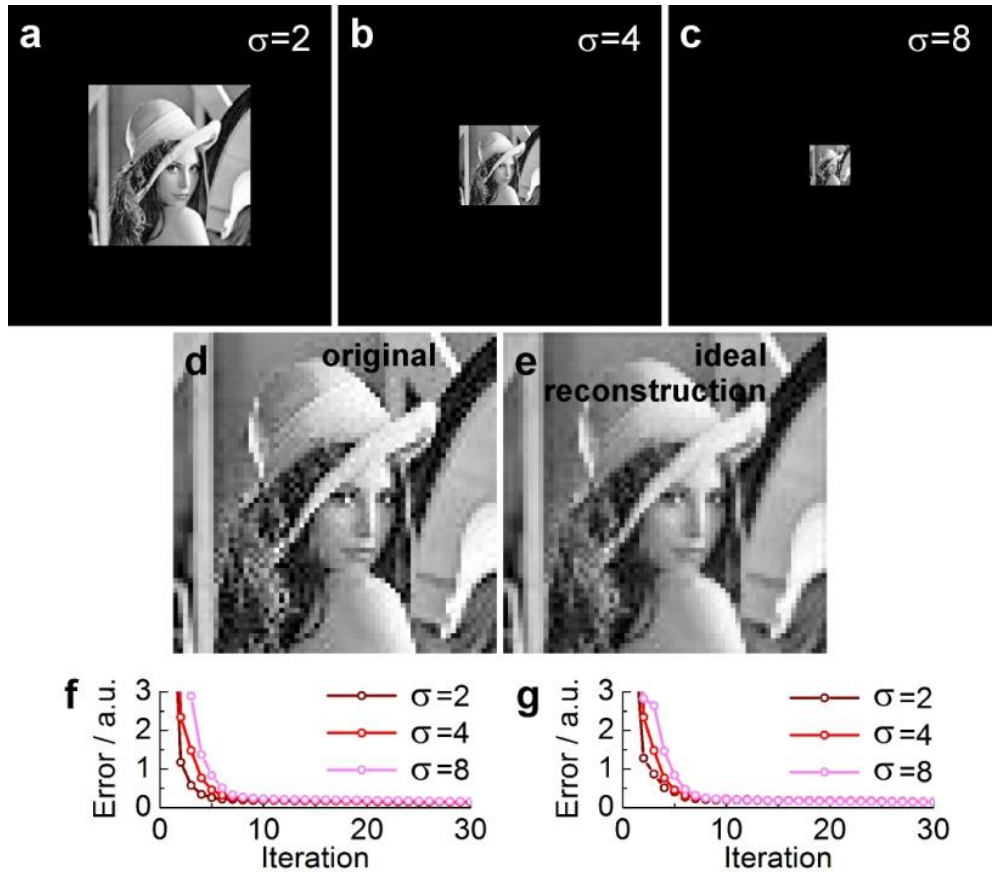

Fig. 4. Effects of oversampling ratio.

(a) – (c) Illustration of different oversampling ratios, the total number of pixels is 256 × 256 pixels and the object size is (a) 128 × 128 pixel ($\sigma$ = 2), (b) 64 × 64 pixel ($\sigma$ = 4) and (c) 32 × 32 pixel ($\sigma$ = 8).

(d) Original object sampled with 64 × 64 pixels and (e) reconstruction obtained from its diffraction pattern. The diffraction pattern was simulated and reconstructed as described in Appendix 1.

(f) and (g) error as a function of the iteration number at different oversampling ratios. (f) The object size is constant and amounts to 64 × 64 pixels; the total sample area size is 128 × 128 pixels ($\sigma$ = 2), 256 × 256 pixels ($\sigma$ = 4) and 512 × 512 pixels ($\sigma$ = 8). (g) The total size remains constant and amounts to 256 × 256 pixels; the object size is 128 × 128 pixels and ($\sigma$ = 2), 64 × 64 pixels ($\sigma$ = 4) and 32 × 32 pixels ($\sigma$ = 8). The parameters of the reconstruction procedure are provided in Appendix 1. The shown error curves are result of averaging over the ten error curves corresponding to the ten selected reconstructions with the smallest errors.

### 3.1.3 Intensity dynamic range of detecting system

A physical detecting system has a finite intensity dynamic range. Figure 5a – d demonstrate the effect of a finite intensity dynamic range of the detector onto the obtained reconstructions: the quality and the resolution of the obtained reconstruction worsens when the number of grey levels decreases. The reason is as follows. When all intensity values in a diffraction pattern are sampled with a finite number of grey levels, the signal at high scattering vectors (at the rim of the diffraction pattern) can be too weak to make even one gray level and therefore it will be recorded as zero. However, the resolution of the obtained reconstruction, in turn, is provided by exactly that signal detected at large scattering angles. When this signal turns to zero due to finite intensity sampling, Fig. 5a – b, high-resolution features become unresolved in the reconstruction, Fig. 5c – d.

A reconstruction obtained from 20 bit grey levels diffraction pattern is somewhat similar to the reconstruction obtained from the original diffraction pattern, compare Fig. 5c and Fig. 4e. When the diffraction pattern intensity is sampled with 16 bit (65536) gray levels, the reconstruction is of a much poorer quality than the reconstruction obtained from the original diffraction pattern, compare Fig. 5d and Fig. 4e.

In practice, to increase the intensity dynamic range, one can measure several diffraction patterns at different exposures and then recombine them into one high dynamic range diffraction pattern [13, 16] by applying available numerical procedures [17].

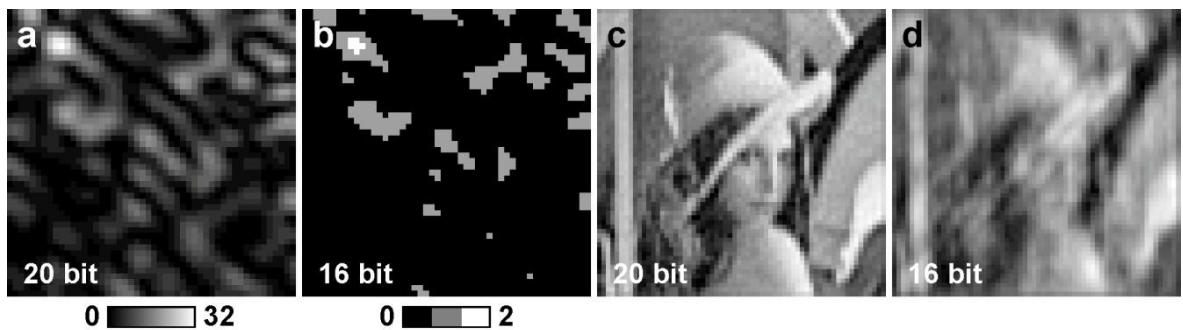

Fig. 5. Effects of the intensity dynamic range of the detector. (a) and (b) Fragments of diffraction patterns when the detecting system has the intensity dynamic range of (a) 20 bit and (b) 16 bit; 50 × 50 pixel left bottom fragments of 256 × 256 pixel diffraction patterns are shown, the intensity scalebar in grey levels. (c) and (d) the corresponding reconstructions obtained from (c) 20 bit and (d) 16 bit intensity range diffraction patterns. The parameters of the diffraction pattern and the reconstruction procedure are provided in Appendix 1.

### 3.1.4 Distribution of probing wavefront

Another factor which can affect the quality of the obtained reconstruction is the distribution of the incident wavefront. For a perfect reconstruction, the incident wavefront should be approximately constant. In reality, however, it is not constant though can be approximated by a Gaussian distribution. The far-field distribution of the scattered wavefront is given by the FT of the product of the object function and the incident wavefront distribution, and can be represented as a convolution of the FT of the object function distribution with the FT of the incident wavefront distribution. Therefore, when the incident wavefront exhibits a Gaussian-distributed amplitude, it acts as a low-pass filter, thus blurring the entire diffraction pattern, including the high-resolution information stored in the fine speckle pattern at large scattering vectors (large $K$-values), as shown in Fig. 6a – b. As a result of this noise re-distribution, the obtained reconstruction does not exhibit high-resolution features as can be seen from the reconstruction shown in Fig. 6c when comparing it to the reconstruction shown in Fig. 4e.

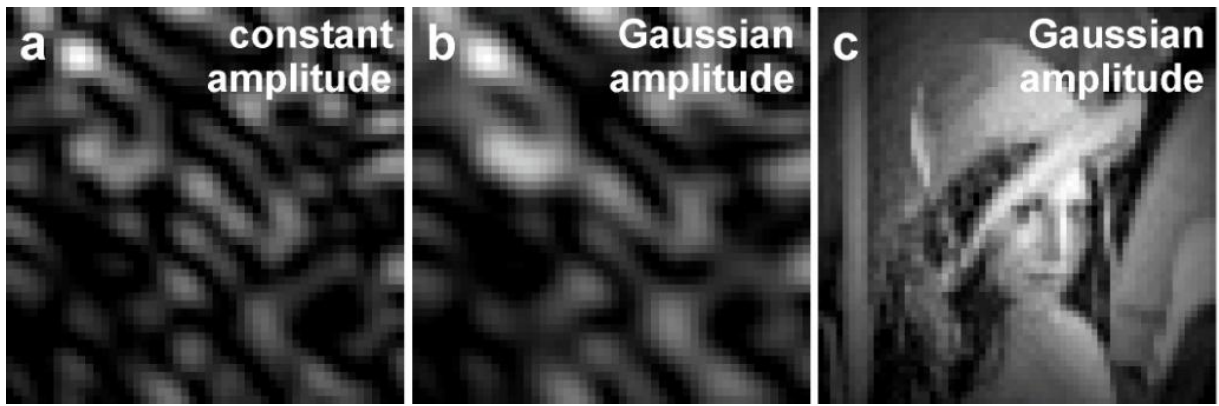

Fig. 6. Effect of the intensity distribution of the incident wavefront on the reconstructed sample distribution. (a) and (b) intensity distributions of the diffraction patterns simulated with an incident wavefront of (a) a constant amplitude and (b) amplitude in form of a Gaussian distribution with standard deviation of 20 pixels; 50 × 50 pixel left bottom fragments of 256 × 256 pixel diffraction patterns are shown. (c) Reconstruction obtained from the diffraction pattern simulated with the incident wavefront with a Gaussian-distributed amplitude. The parameters of the diffraction pattern and the reconstruction procedure are provided in Appendix 1.

**3.1.5 Effect of a noise in diffraction pattern**

The Fourier transform of white noise is still noise. Therefore, when noise is added to the distribution in the Fourier domain, one might expect, as a consequence, noise to be added to the corresponding real-space distribution. However, in CDI when noise is added to the intensity (diffraction pattern) distribution, the consequence is not just noise being added to the reconstruction. The reason is that during the iterative phase retrieval, all values outside the object support (which also could be associated with noise) are forced to be zero. As a result, the quality and the resolution of the reconstruction worsen.

Figure 7 shows the effect of noise added to the diffraction pattern, Fig. 7a – b, and its effect onto the reconstruction, Fig. 7c – e. At a signal-to-noise ratio SNR = 5 (Fig. 7e), the quality of single reconstruction with the least error is so poor that the original object can hardly be recognized. Only after averaging over 10 reconstructions with the smallest errors (Fig. 7d) does the reconstructed distribution resembles the original object. Martin et al suggested "noise tolerant HIO" algorithm which allows reconstructing a recognizable object from a single noisy diffraction pattern [18].

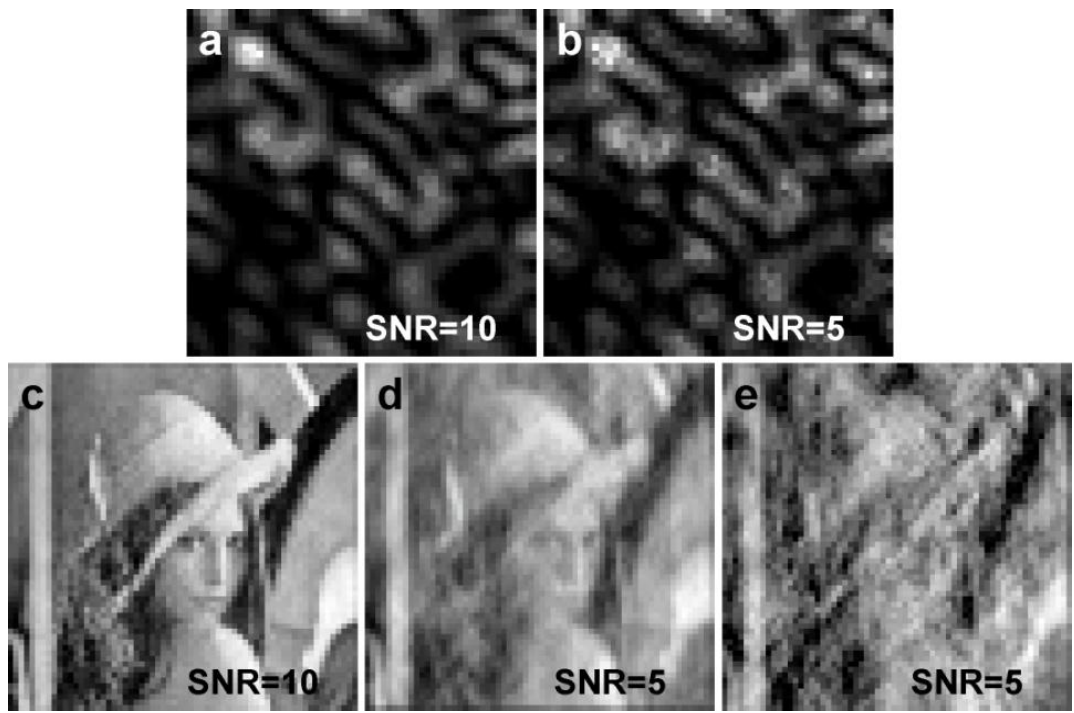

Fig. 7. Effect of noise in diffraction pattern. (a) – (b) 50 × 50 pixel left bottom fragments of 256 × 256 pixel diffraction patterns with signal-to-noise ratios (SNR) of (a) SNR = 10 and (b) SNR = 5 are shown. (c) – (d) Reconstructions obtained from diffraction pattern with (c) SNR = 10 and (d) SNR = 5. (e) Reconstruction with the least error obtained from diffraction pattern with SNR = 5. The parameters of the diffraction patterns and the reconstruction procedure are provided in Appendix 1.

**3.1.6 Distortions in diffraction pattern**

An experimental diffraction pattern can have distortions when compared to an ideal diffraction pattern (Fig. 8). Distortions can occur due to misalignment in the optical system or a non-flat detector surface. For electron imaging, distortions can also be caused by the deflection of electrons by residual electromagnetic fields in the system. The effect of such distortions on reconstruction is shown in Fig. 8. In the case when the diffracted wave is detected at different $Z$-distances (Fig. 8b and c), the resultant reconstruction still resembles that obtained from an ideal diffraction pattern (Fig. 8d). The effect of lateral distortions in the diffraction pattern is much more severe: displacements by a factor 1.05 in the $X$ and $Y$ directions in one-quarter of the diffraction pattern (as illustrated in Fig. 8e) result in a reconstruction which hardly resembles the original distribution (Fig. 8f).

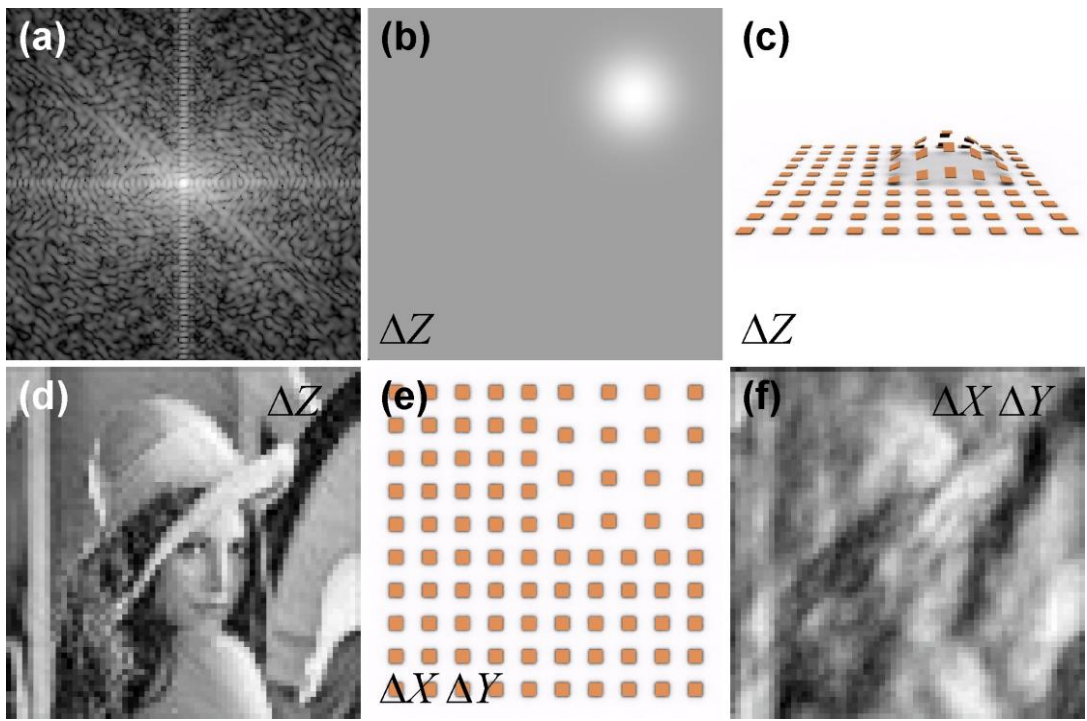

Fig. 8. Effect of distortions in diffraction pattern. (a) Original diffraction pattern. (b) and (c) illustrations to $Z$-distortion in the upper right quadrant of the diffraction pattern: the variation in the $Z$-distance of the detecting system ranges from 0.1 m to 0.11 m. (d) The corresponding reconstruction obtained from the diffraction pattern with the $Z$-distortion. (e) Illustration of lateral distortions in the upper right quadrant, not drawn to scale. (f) Reconstruction obtained from the diffraction pattern where the coordinates of the detected intensities are shifted in the upper right quadrant by a factor 1.05 along both $X$- and $Y$-directions. The diffraction pattern is simulated for a wavelength of 500 nm, $Z$ = 0.1 m, and sample area size of 1 mm × 1 mm. The parameters of the reconstruction procedure are described in Appendix 1.

## 3.2 Data treatment before phase retrieval

Any experimental diffraction pattern needs to be prepared before it can be put into an iterative phase retrieval routine. In this section, we consider the issues that should be taken care of during the preparation of the experimental diffraction pattern for iterative reconstruction.

### 3.2.1 Centering of diffraction pattern

Diffraction pattern should be centered such that the intensity maximum is in the center. Even a small offset by one pixel can have a dramatic effect on the resultant reconstruction, as illustrated in Fig. 9. The effect of diffraction pattern off-centering depends on the oversampling ratio, the higher the oversampling ratio the less is the effect. At low oversampling ratio ($\sigma$ = 4) the reconstructions obtained from one-pixel shifted diffraction pattern are significantly worse than the original reconstruction (Fig. 9a and b). At higher oversampling ratio ($\sigma$ = 8) the reconstruction obtained from one-pixel shifted diffraction pattern is almost as good as the original reconstruction (Fig. 9c), and thus the effect of diffraction pattern off-centering is negligible. If the information in the center of the diffraction pattern is missing (due to beamstop or overexposure), then the diffraction pattern can be centered by comparing the distribution of higher-order centro-symmetric peaks (or speckles).

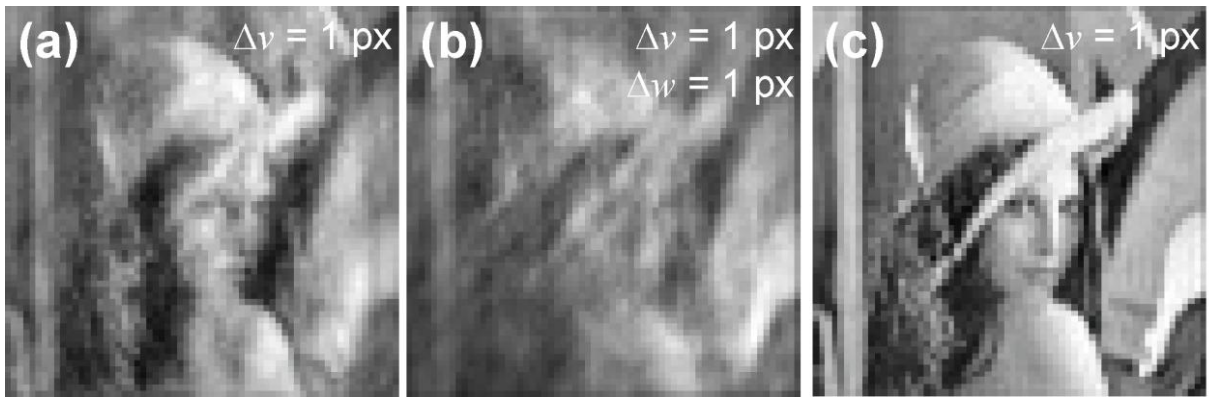

Fig. 9. Effect of centring of diffraction pattern. Reconstructions obtained from a diffraction pattern with oversampling ratio $\sigma$ = 4, which was off-centered by (a) 1 pixel in the *v*-direction and (b) 1 pixel in both v- and *w*-directions. (c) Reconstructions obtained from a diffraction pattern with oversampling ratio $\sigma$ = 8, which was mis-centered by 1 pixel in *v*-direction. The parameters of the reconstruction procedure are described in Appendix 1.

### 3.2.2 Effect of additional constant background signal

For successful reconstruction, the intensity distribution of the diffraction pattern should not contain a constant off-set, as demonstrated in Fig. 10. A small additional constant background of 1×10⁻⁶ of the maximal intensity in the diffraction pattern does not seriously affect the quality of the reconstruction, Fig. 10a. However, as the additional constant background increases, the quality of reconstruction worsens and a strong signal appears at single pixels in the center, as shown in Fig. 10b and c.

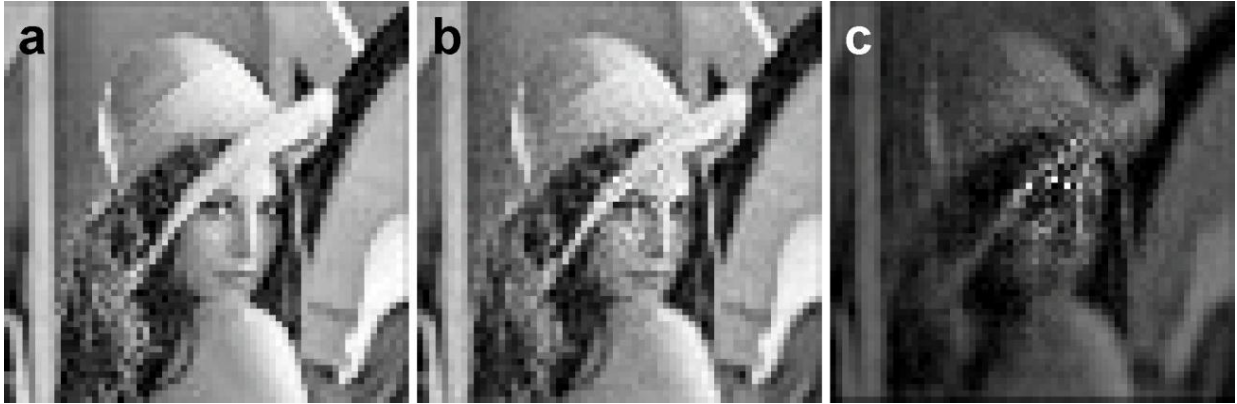

Fig. 10. Effect of a constant background added to diffraction pattern. Reconstructions obtained from the diffraction pattern when a constant of (c) 10⁻⁶, (d) 10⁻⁵ and (e) 10⁻⁴ of the diffraction pattern maximal intensity is added to the diffraction pattern. The parameters of the diffraction pattern and the reconstruction procedure are provided in Appendix 1.

### 3.2.4 Real-valued objects and symmetrization of diffraction pattern

A priori information about the object can be extremely useful for optimizing the input data and the constraints for the phase retrieval. For example, if it is known that the object is real-valued, it means that its diffraction pattern must be centro-symmetric. If the acquired diffraction pattern is not centro-symmetric, it is most likely due to the noise, and the diffraction pattern can be symmetrized by applying the following procedure:

$$I_{\text{sym}}(v,w) = \frac{I(v,w) + I(-v,-w)}{2}.$$

If the diffraction pattern is not symmetrised, its asymmetry will lead to non-zero values of the phase distribution in the reconstructed object distribution, which at each iteration will get into conflict with the constraint that the object should be real-valued. For this reason, symmetrization of the diffraction pattern can be applied to ensure the stability of the reconstruction process.

## 3. 3 Iterative phase retrieval

### 3.3.1 Object support

For many phase retrieval algorithms the knowledge of the exact object support (the extent of the area occupied by the object) is not required and setting the object support to a square patch is sufficient [1, 19]. When necessary, the object support can be obtained by taking an image of the sample by some other imaging technique. Also, the object support can be evaluated from the object auto-correlation function, which is in turn calculated by taking the FT of the diffraction pattern. The extent of the auto-correlation function distribution corresponds to twice the extent of the object, as illustrated in Fig. 11.

The most effective way to recover the sample support is by applying the shrinkwrap algorithm, which estimates the sample support at the first iteration (from the auto-correlation function by FT of the diffraction pattern) and then iteratively recovers the exact object shape by periodically updating the object support during the iterative procedure [9].

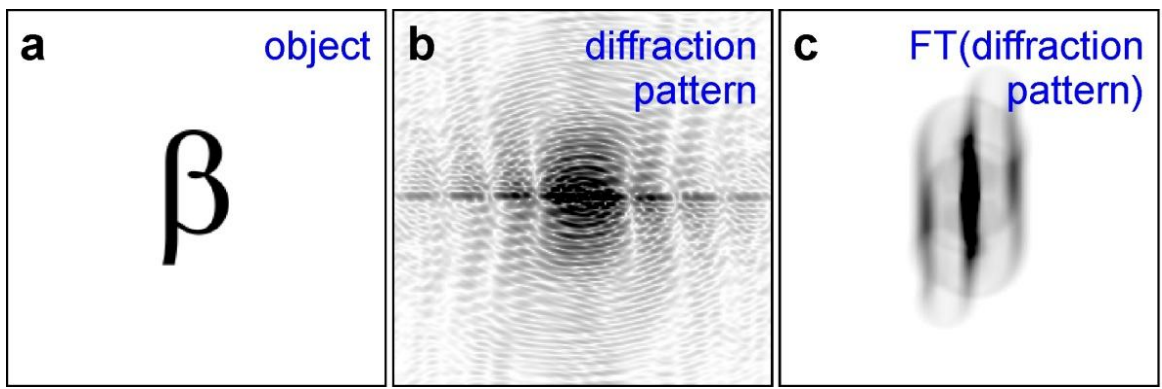

Fig. 11. Auto-correlation of the object. (a) Original object and (b) its diffraction pattern shown in an inverted logarithmic intensity scale. (c) Amplitude of the Fourier transform of the diffraction pattern shown in (b). The obtained distribution consists of four centro-symmetric images of the original object.

### 3.3.2 Pixel with artifact values in diffraction pattern

Detectors often have some faulty pixels, which deliver incorrect values. For example, it can be "dead" pixels with zero intensity value or saturated ("bright") pixels with extremely high intensity value. If these values are kept during the iterative phase retrieval procedure, the resultant reconstruction will be marred by artifacts as illustrated in Fig. 12. Obviously, a perfect detector is the best solution. A practical solution to the problem of signals originating from pixels with incorrect values is (i) to identify all the pixels with missing or wrong signals before starting the iterative phase retrieval, and (ii) during the iterative phase retrieval (as expressed by Eq. 10 and 11) replace the values of those pixels with the iterated amplitudes $|G_k(v,w)|$ [20]. This way, the correct values at the pixel with the missing or wrong signal are recovered, and simultaneously an artifact-free reconstruction is achieved, compare Fig. 12d and h to Fig. 4e.

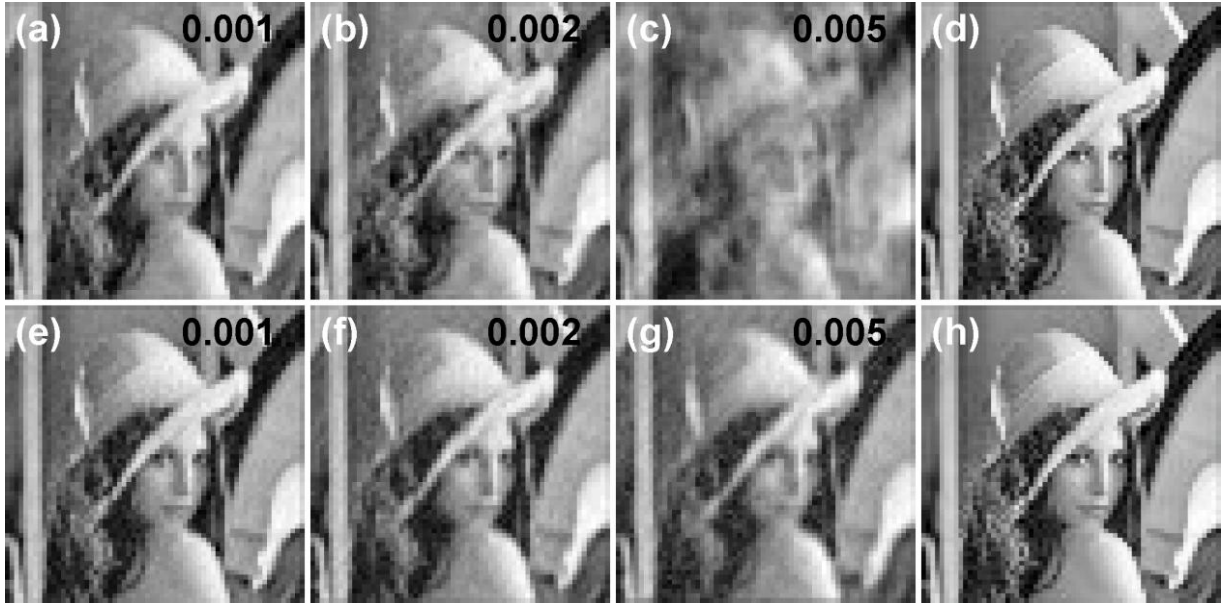

Fig. 12. Effect of pixels with artifact values in diffraction pattern. To mimic the experimental situation, the values of pixels at randomly distributed positions are set to wrong values. (a) - (d) Effect of "dead" pixels with zero values. (a) - (c) Reconstructions obtained from diffraction patterns where the number of pixels with values of zero is (a) 0.001, (b) 0.002 and (c) 0.005 of the total number of pixels. (d) Reconstruction obtained by the HIO algorithm where at each iteration, the values at the pixels with missing values were replaced by the iterated amplitudes $|G_k(v,w)|$. (e) – (h) Effect of saturated ("bright") pixels with values of 0.1 of the intensity maximum in the diffraction pattern. (e) - (g) Reconstructions obtained from diffraction patterns where the number of pixels with artifact values is (e) 0.001, (f) 0.002 and (g) 0.005 of the total number of pixels. (h) Reconstruction obtained by the HIO algorithm where at each iteration, the pixels with artifact values were replaced by the iterated amplitudes $|G_k(v,w)|$. The parameters of the diffraction patterns and the reconstruction procedure are described in Appendix 1.

### 3.3.3 Missing central spot

In a typical high-resolution x-ray diffraction experiment, the central part of the diffraction pattern is not available due to either a hole in the detector, a beamstop, or a saturated (overexposed by direct beam) region. It is excluded to avoid the direct beam and to acquire the remaining diffraction pattern, including the higher-order scattered signal, at high dynamic range. The missing central part contains the low-resolution information about the overall shape of the object. For example, it can be replaced by the squared amplitude of the FT of image of the sample image, which was acquired by a low-resolution imaging technique [1, 19, 21]. Numerically, the missing central spot can be recovered as described above; the values of the missing central spot pixels are replaced with the iterated amplitudes $|G_k(v,w)|$ [20] during the iterative phase retrieval (as expressed by Eq. 10 and 11).

### 3.3.4 Recovery of phase objects

Already at the beginning of the development of phase retrieval algorithms, it was demonstrated that phase objects are more difficult for reconstruction than amplitude-only objects; however, they can be successfully recovered, provided their exact support is known [22, 23]. It was later demonstrated that a phase object can be reconstructed without a priori knowledge about its support; for example, by applying the shrinkwrap algorithm [9] or by applying the HIO algorithm with a loose support and additional object phase constraint that allows object phase values within a certain range [24]. Although reconstruction of complex-valued objects require modified algorithms [10], similar effects of diffraction pattern properties (noise, off-set, oversampling ratio, etc) on the resultant iterative phase retrieval reconstruction are also expected in the case of complex-valued object. The simulations provided in this study are therefore limited to the case of real-valued objects.

### 3.3.5 Recovery of binary objects

Binary objects are those which have transmission function described by a binary function (0 - no transmission, 1 - all signal transmitted) and often these are referred to as "mask" objects [1, 19, 20]. Although it seems that such objects are easier to reconstruct, this is not entirely correct because: (1) the convergence and the error of the phase retrieval are no different from those in the case of a non-binary object (as illustrated in Fig. 13), and (2) there are no binary objects in reality – any mask has no sharp edges and hence no binary transmission.

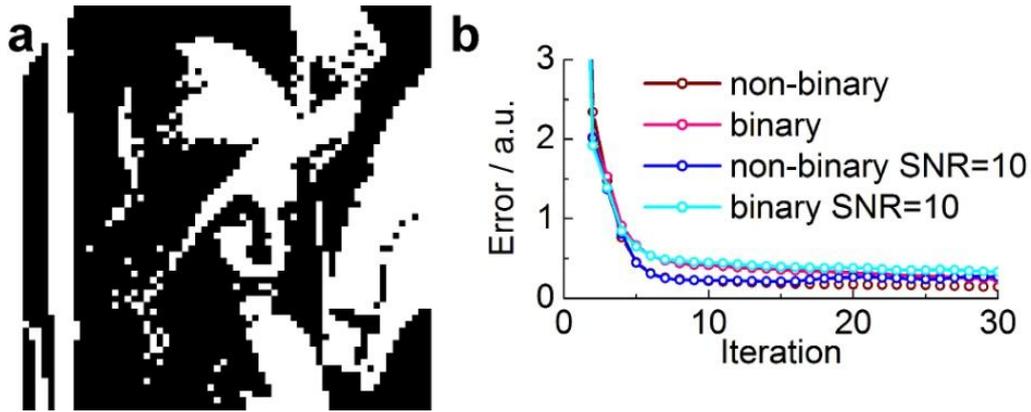

Fig. 13. Recovery of binary objects. (a) The binary object created from "Lena" image. (b) Error as a function of iteration number for diffraction patterns of the non-binary and binary "Lena" image, noise-free and with SNR = 10. The shown error curves are the result of averaging over the ten error curves corresponding to the ten selected reconstructions with the smallest errors. The parameters of the diffraction pattern and the reconstruction procedure are provided in Appendix 1.

## 3.4 Optimized phase retrieval algorithms

Most of the phase retrieval algorithms are based on the Gerchberg-Saxton algorithm [8]: ER [2], HIO [2], shrinkwrap [9], charge-flipping algorithm [25], relaxed averaged alternating reflections (RAAR) algorithm [26], noise tolerant HIO algorithm [18], and many others [10]. The main differences between these algorithms are the various constraints applied in real space [2, 9, 20, 24] and reciprocal space [27-29], which can be optimized depending on the particular sample and experiment. Moreover, a combination of phase retrieval algorithms (ER and HIO algorithms [2, 24, 30]) is often applied in an alternating fashion to avoid stagnation or oscillation of the iterative process and to stabilize the final reconstruction.

As a preliminary reconstruction, a low-resolution reconstruction can be obtained by selecting only the central part of the diffraction pattern, as demonstrated in Fig. 14. The oversampling condition translates into a certain size of the pixel in the detector plane, and once the oversampling condition is fulfilled, it is also fulfilled for a cropped version of the same diffraction pattern. If the diffraction pattern is cropped from N × N pixels to $N_0 \times N_0$ pixels, then the reconstruction obtained from the cropped diffraction pattern will correspond to the same area as the reconstruction obtained from the original non-cropped diffraction pattern but sampled with $N_0 \times N_0$ pixels, as shown in Fig. 14. The advantage of obtaining a low-resolution reconstruction is that a diffraction pattern sampled with a smaller number of pixels can be reconstructed much faster.

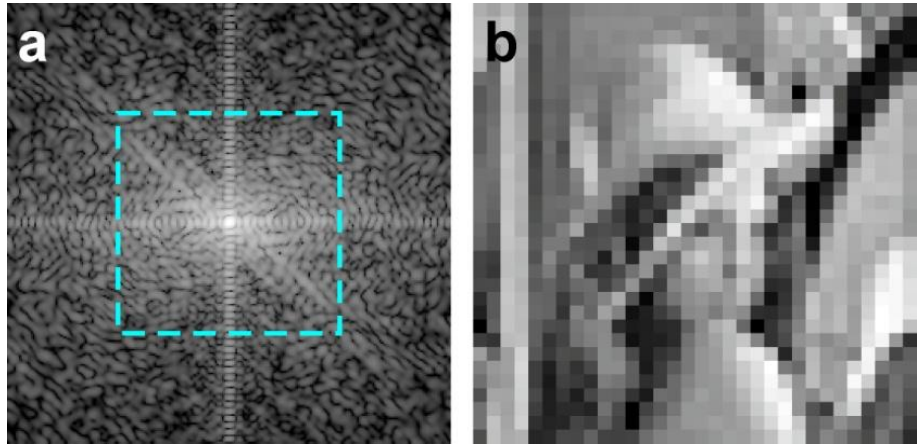

Fig. 14. Obtaining low-resolution reconstruction. (a) Original 256 × 256 pixels diffraction, oversampling ratio $\sigma$ = 4. (b) Reconstruction obtained from the central part of 128 × 128 pixel of the original diffraction pattern as indicated by the dashed square in (a). The obtained reconstruction has a size of 32 × 32 pixels. The reconstruction was obtained by applying the HIO algorithm with the tight support in the form of a squared patch of 32 × 32 pixels and real and positivity constraint for 2000 iterations; 10 reconstructions with the smallest errors were selected, aligned and averaged.

### 3.5 Goodness of reconstruction and PRTF

As already mentioned above, the goodness of the final reconstruction can be evaluated by error function, which can be calculated by Eq. 12 and 13. Typically, one single reconstruction with the least error is not representative, and a better reconstruction can be achieved if many reconstructions are obtained and several reconstructions with the smallest errors are selected and averaged (an example is shown in Fig. 7d and e). For example, 100 reconstructions are obtained and 10 reconstructions with the smallest errors are selected, aligned and averaged. The reconstructed object can appear at any lateral position in the object plane (unless a fixed tight support was applied in the iterative routine). Also, for a real-valued object, the distribution of the reconstructed object can be centro-symmetrically flipped. Therefore, the reconstructed objects have to be accordingly flipped and aligned prior to their summation. The alignment procedure (also called as "image registration") can be done by applying a cross-correlation approach that is common in sub-pixel image registration methods [31]. The resultant average reconstruction exhibits a better SNR than one reconstruction with the least error, compare the reconstructions shown in Fig. 7d and e.

Because the spectrum of a finite object is infinite, and a diffraction pattern measures only a fraction of that spectrum, any recovered object distribution is only *an approximate* solution to the true object distribution. Also, the recovered complex-valued distribution of the scattered wave in the far field is only an approximate solution to the true complex-valued distribution. The parameter that calculates how well the recovered amplitudes match the original amplitudes is the phase retrieval transfer function (PRTF) [32, 33]:

$$\text{PRTF}(\vec{u}) = \frac{|\langle \Gamma(\vec{u}) \rangle|}{|F(\vec{u})|},$$

where $|F(\vec{u})|$ are the measured amplitudes, $\Gamma(\vec{u})$ are the iteratively recovered amplitudes (i. e. FT of the object reconstruction), $\vec{u} = (v, w)$ is the coordinate in the Fourier domain, and $\langle ... \rangle$ denotes averaging over several reconstructions. $\langle \Gamma(\vec{u}) \rangle$ is calculated by applying the following relations:

$$\langle \Gamma(\vec{u}) \rangle = \frac{1}{M} \sum_{m=1}^{M} \Gamma_m(\vec{u}) = \frac{1}{M} \sum_{m=1}^{M} \text{FT}[g_m(x,y)] = \frac{1}{M} \text{FT}\left[\sum_{m=1}^{M} g_m(x,y)\right].$$

where $g_m(x,y)$, $m=1...M$ are the aligned object reconstructions with the smallest errors.

The PRTF compares the recovered complex-valued amplitude distributions to the measured amplitude distribution. If the recovered phases are consistently reconstructed, then $\langle \Gamma(\vec{u}) \rangle$ gives the complex-valued distribution with the amplitude distribution close to $|F(\vec{u})|$ and thus PRTF is close to unity. If the recovered phases are random, then $\langle \Gamma(\vec{u}) \rangle$, and therefore the PRTF, are both close to zero. For a noise-free diffraction pattern, PRTF is close to unity; as the noise in a diffraction pattern increases, the PRTF exhibits more and more decay at higher frequencies (plots of the PRTFs, averaged over circles of constant $u$ are shown in Fig. 15). The PRTF is also employed as a resolution criterion. The recovered resolution is defined by the point where the PRTF drops to 1/e [34] or to 0.5 [33, 35]. It should be noted that the overall applicability of PRTF often causes discussions [36].

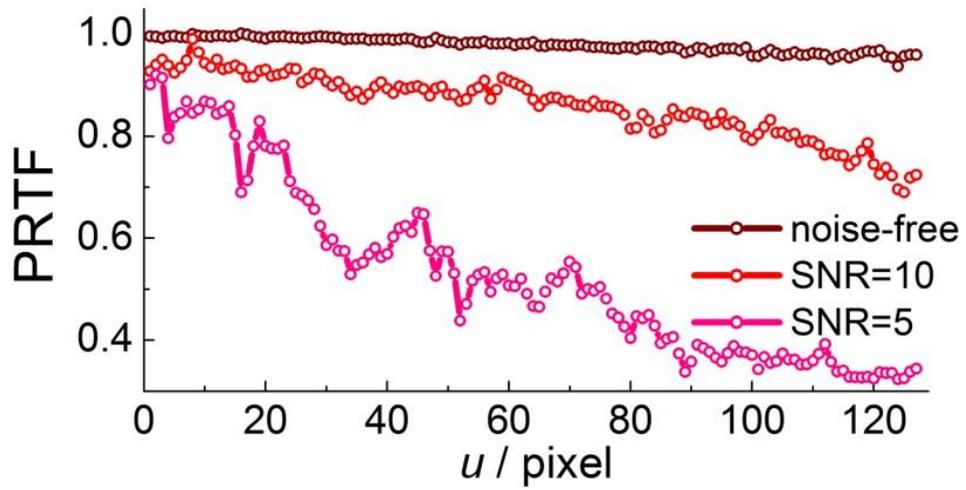

Fig. 15. Phase retrieval transfer function (PRTF) for a noise-free diffraction pattern and for diffraction patterns with SNR = 10 and SNR = 5. The parameters of the diffraction patterns and the reconstruction procedure are provided in Appendix 1.

## 3.6 Simulation of diffraction patterns

### 3.6.1 Simulation without fast Fourier transforms

All simulated diffraction patterns shown in this article were calculated without application of fast Fourier transforms (FFT) to avoid the wrapping of signal at the rim of images. The simulation procedure was as follows. The sample distribution was digitized, that is, it was represented in pixels. For each pixel, the diffracted complex-valued wavefront was calculated as the analytical solution of diffraction on a square aperture. The total sum of complex-valued wavefronts from all pixels yielded total diffracted wavefront. The squared amplitude of the total diffracted wavefront gave the intensity distribution of the diffraction pattern.

### 3.6.2 Wrapping effect

Wrapping of a signal occurs when FFT is applied. When the Fourier transform of a finite object distribution is calculated analytically, the obtained spectrum is infinite. When the Fourier transform is calculated numerically via FFT, the range of the obtained spectrum is finite and given by the pixel size in the real space: $2K_{max} = \frac{1}{N\Delta_x}$, where $\Delta_x$ is the pixel size in the real space. However, all frequencies of the signal spectrum still show up in the calculated spectrum, but some of them show up at wrong positions. Low frequencies of the signal spectrum, that are at $|K| < K_{max}$ are correctly represented. Higher frequencies that are beyond the provided spectrum range, which are at $|K| > K_{max}$, are "wrapped" around the edges of the spectrum range, or "reflected" from the spectrum edge back into the spectrum finite range, as illustrated in Fig. 16.

It can be speculated that the effect of signal wrapping at the edges can also affect reconstructions obtained by iterative phase retrieval. It is however impossible to check it by simulations, because it would require an iterative routine which does not utilize FFTs and employs analytical solutions of non-digitized signals. For a typical experimental diffraction pattern, the signal at the edges is close to zero, and thus the signal wrapping at the edges should not have a significant effect on the resultant reconstruction. When the signal at the edges is relatively intense, then the cropped Fourier spectrum will have a more significant effect on the resultant reconstruction (Gibbs phenomenon) than the wrapping of the signal at the edges. It can be also speculated that signal wrapping at the edges can be one of the reasons why PRTF decays at higher frequencies even for an ideal noise-free simulations.

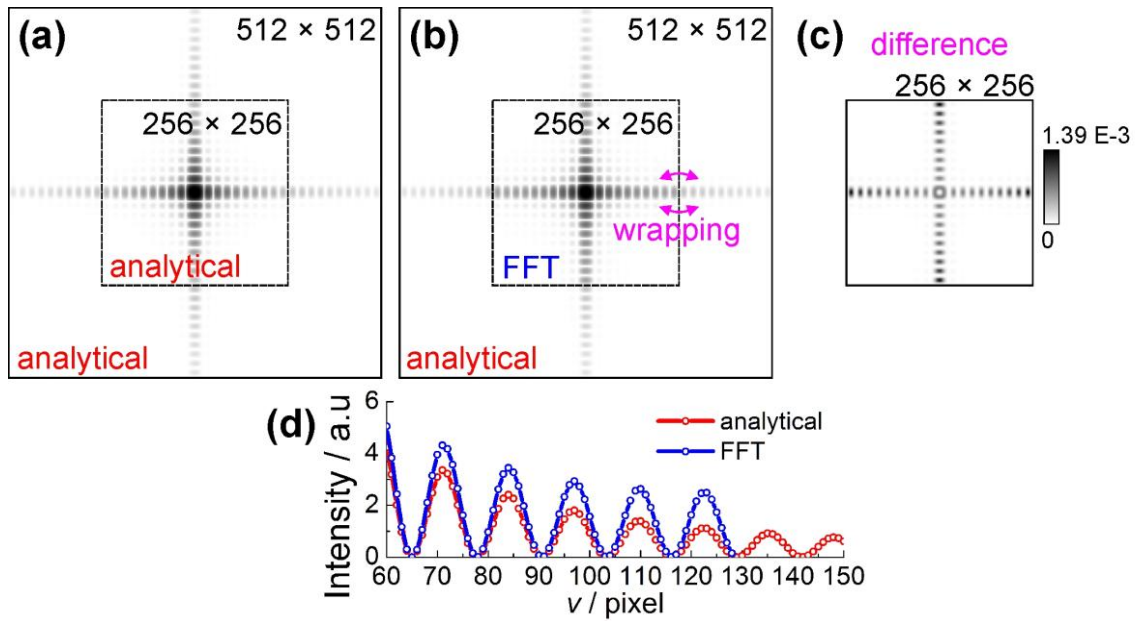

Fig. 16. Wrapping effect. Diffraction on a square aperture is considered. (a) Diffraction pattern calculated as an analytical solution to the problem of diffraction on a square aperture, calculated for 256 × 256 pixels (DP1) and 512 × 512 pixels (DP2). (b) Diffraction on a square aperture calculated by FFT for 256 × 256 pixels (DP3). The spectrum which exceeds region 256 × 256 pixels is wrapped around the edges and shows up inside the 256 × 256 pixels window. (c) The difference between the diffraction patterns simulated as an analytical solution and by FFT, calculated as (DP3-DP1)/max(DP1). It is apparent that the difference is maximal at the edges. (d) Intensity profiles through the center of diffraction patterns shown in (b), demonstrating the difference in intensity values at the coordinate of $v = 128$ pixel (wrapping).

### 3.6.3 One pixel problem

The following notion should be kept in mind when performing a simulation of a far-field distribution. One pixel has a finite size. When calculating wavefront scattered by an object, which has the size of one pixel, the analytical solution of the far-field distribution of the scattered wave corresponds to diffraction on a square aperture, which will be represented by a sinc function. However, when the scattered wave is calculated by FFT of a distribution of all zeroes expect one pixel (which has a value of 1), the resultant distribution will be the FFT of a delta function and the solution will be represented by a constant. This mismatch between real space sizes and digital representation must be taken into account when performing simulations.

# 4. Discussion and conclusions

In conclusion, we briefly summarize which parameters are crucial and which are not for the successful iterative phase retrieval. Most importantly, the experimental parameters must be selected in such a way that the oversampling condition is fulfilled. The magnitude of the oversampling factor is a less crucial factor. A high intensity dynamics range and absence of distortions in the acquired diffraction pattern are necessary for successful reconstruction. Faulty pixels of the detecting system or missing central spot are less of a problem as they can be recovered during the iterative phase retrieval. Conventional algorithms like ER, HIO, shrinkwrap and/or their combination can be adjusted and optimized for particular experimental data. As we showed here, a preliminary low-resolution reconstruction can be quickly obtained from a cropped diffraction pattern which can help to refine settings for full-resolution reconstruction from the full-size diffraction pattern.

Appendix 1

All diffraction patterns shown here, unless other is specified, were simulated using the following parameters: object ("Lena" image) was sampled with 64 × 64 pixels and zero-padded to 256 × 256 pixels, which gives an oversampling ratio $\sigma$ = 4. One hundred reconstructions were obtained by applying th eHIO algorithm with tight object support in form of a square patch of 64 × 64 pixels and the constraint that the object must be real and positive; a total of 2000 iterations were made. 10 reconstructions with the smallest errors (calculated with Eq. 13) were selected, aligned and averaged.